\documentclass[amsmath,amssymb,twocolumn,superscriptaddress]{revtex4}

\usepackage{graphicx}%
\usepackage{dcolumn}%
\def\be{\begin{equation}}
\def\ee{\end{equation}}
\def\beq{\begin{eqnarray}}
\def\eeq{\end{eqnarray}}
\def\n{\nonumber}
\usepackage{bm}
\usepackage{graphicx}
\usepackage{dcolumn}
\usepackage{amsmath}
\usepackage[latin1]{inputenc}
\usepackage{graphicx, psfrag}
\usepackage{amssymb}
\usepackage[colorlinks=true, citecolor=blue, urlcolor = blue, linkcolor= red, bookmarks=true]{hyperref}
\usepackage{float}
\usepackage{amsmath}
\usepackage{amsfonts}
\usepackage{dcolumn}
\usepackage{hyperref}
\usepackage{subfigure}
\usepackage{pgfplots}
\usepackage{epstopdf}
\usepackage{booktabs}
\usepackage{orcidlink}
\usepackage{xcolor}

\begin{document}


\title{Quantum corrections to Dymnikova-Schwinger black holes in Einstein-Gauss-Bonnet gravity} 

\author{A. Errehymy\orcidlink{0000-0002-0253-3578}}
\email[]{abdelghani.errehymy@gmail.com}
\affiliation{Astrophysics Research Centre, School of Mathematics, Statistics and Computer Science, University of KwaZulu-Natal, Private Bag X54001, Durban 4000, South Africa}
\affiliation{Center for Theoretical Physics, Khazar University, 41 Mehseti Str., Baku, AZ1096, Azerbaijan}

\author{Y. Khedif}
\email[]{youssef.khedif@gmail.com}
\affiliation{Laboratory of High Energy and Condensed Matter Physics, Department of Physics, Faculty of Sciences A\"in Chock, University Hassan II, P.O. Box 5366 Maarif Casablanca 20100, Morocco}

\author{M. Daoud}
\email[]{m$_{}$daoud@hotmail.com}
\affiliation{Department of Physics, Faculty of Sciences, Ibn Tofail University, P.O. Box 133, Kenitra 14000, Morocco}
\affiliation{Abdus Salam International Centre for Theoretical Physics, Miramare, Trieste 34151, Italy}

\author{K. Myrzakulov}
\email[]{krmyrzakulov@gmail.com}
\affiliation{Department of General and Theoretical Physics, L.N. Gumilyov Eurasian National University, Astana 010008, Kazakhstan }

\author{B. Turimov}
\email[]{bturimov@astrin.uz}
\affiliation{Alfraganus University, Yukori Karakamish Str. 2a, 100190 Tashkent, Uzbekistan}
\affiliation{University of Tashkent for Applied Sciences, Gavhar Str. 1, Tashkent, 100149, Uzbekiston}
\affiliation{Shahrisabz State Pedagogical Institute, Shahrisabz Str. 10, Shahrisabz, 181301,Uzbekiston}

\author{T. Myrzakul}
\email[]{tmyrzakul@gmail.com}
\affiliation{Department of Computer Science, L.N. Gumilyov Eurasian National University, Astana, 010008, Kazakhstan }

\begin{abstract}
This work investigates black holes within a modified framework of gravity that incorporates quantum-inspired corrections and a fundamental minimal length scale. By integrating Einstein-Gauss-Bonnet gravity with a specially tailored matter source that models quantum particle creation, we derive novel, non-singular black hole solutions. These black holes exhibit rich horizon structures and, notably, do not undergo complete evaporation---instead, they stabilize into permanent remnants. In addition to analyzing the thermodynamic implications of quantum corrections to Dymnikova-Schwinger black holes, we examine their quasinormal mode spectra using the WKB approximation, alongside their associated energy emission rates. Our findings provide compelling new perspectives on how quantum effects may address foundational issues such as the black hole information loss paradox.
\end{abstract}

\maketitle

\date{\today}

\textbf{Introduction:} Black holes have captured the imagination of scientists and astronomers for generations. These mysterious cosmic objects were first proposed over a hundred years ago, thanks to Karl Schwarzschild's work with Einstein's theory of general relativity (GR) \cite{Schwarzschild:1916sf}. Today, we believe black holes can form when massive stars reach the end of their lives and collapse under their own gravity \cite{Oppenheimer:1939ue}. But they're not limited to just stellar remnants--some are unimaginably large, lurking at the hearts of galaxies \cite{Kormendy:2001hb}, while others may have formed in the chaotic moments just after the Big Bang \cite{Suh:2024jbx}. Recent breakthroughs in astronomical imaging, especially with the Event Horizon Telescope, have changed how we capture images of compact astrophysical objects. These advancements provide us with remarkable insights into their properties and behaviors. Two standout examples are the images of the center of the M87 galaxy \cite{EventHorizonTelescope:2019dse} and Sagittarius A*, the heart of our Milky Way \cite{EventHorizonTelescope:2022wkp}. These visuals reveal luminosity profiles of accretion disks that match well with black hole theories, offering strong evidence for their existence and giving us a stunning glimpse into these mysterious objects.

One exciting area of research looks at how alternative theories of gravity stack up against Einstein's GR, especially in the extreme environments around black holes. Instead of relying only on traditional tests, scientists are now using black hole images \cite{Vagnozzi:2022moj} and gravitational wave signals from colliding objects \cite{Mastrogiovanni:2020gua} to explore how gravity really works. These tools could reveal new aspects of physics and help bridge the gap between gravity and quantum mechanics. A fascinating idea in this effort is the Generalized Uncertainty Principle (GUP), which adds a tiny but fundamental length scale to our understanding of black holes \cite{Konishi:1989wk, Amati:1988tn, Amati:1987wq, Gross:1987kza}, possibly pointing the way to a deeper theory of gravity.

Unlike traditional black holes with a central singularity, regular black holes are thought to have a finite-sized core. First proposed by Bardeen \cite{Bardeen:1968}, this idea introduced more realistic models of black hole interiors \cite{Frolov:2016pav}. Some feature de Sitter-like cores \cite{Dymnikova:1992ux}, while others involve a \textit{radial bounce} that reshapes their internal geometry. These structures differ greatly from standard GR predictions, even in Bardeen-type solutions. Research in this area is growing \cite{Menchon:2017qed, Bejarano:2017fgz}, exploring a variety of models---some in higher dimensions, others involving exotic matter, quantum effects, or hybrid forms called \textit{black bounces} \cite{Ayon-Beato:1998hmi, Ansoldi:2006vg, Maluf:2022jjc, Simpson:2019cer}.

Back in the 1990s, Dymnikova introduced a model of a regular black hole \cite{Dymnikova:1992ux} that brought Gliner's idea to life---that a de Sitter-like core could prevent a singularity from forming \cite{Gliner:1966, Gliner:1970}. Later on, researchers \cite{Dymnikova:1996plb, Ansoldi:2008jw} suggested that the density profile in Dymnikova's four-dimensional model might be the gravitational equivalent of the Schwinger effect. Separately, other studies introduced a correction to the Schwinger effect based on the GUP \cite{Haouat:2013yba, Ong:2020tvo}. Inspired by these connections, some of the current authors recently proposed a GUP-based modification to Dymnikova's model and explored its impact on wormhole structures \cite{Estrada:2023pny}. Building on this concept, we explore black holes within a modified gravity framework that incorporates quantum-inspired effects through a fundamental minimal length scale. Our approach combines Einstein-Gauss-Bonnet (EGB) gravity---a natural extension of GR that adds a quadratic correction known as the Gauss-Bonnet (GB) term to the Einstein-Hilbert action---with a matter source modeling quantum particle creation. Originally developed by Lanczos \cite{Lanczos:1938sf} and Lovelock \cite{Lovelock:1971yv, Lovelock:1972vz}, EGB theory emerges as a low-energy limit of string theory and is particularly relevant in higher dimensions. Notably, it remains ghost-free \cite{Boulware:1985wk}, with equations involving only second derivatives of the metric, and is stable when expanded around flat spacetime \cite{Boulware:1985wk, Wiltshire:1985us}. Using this framework, we uncover new, non-singular black hole solutions that feature diverse horizon structures and, importantly, do not fully evaporate-instead leaving behind stable remnants. Our results offer insights into how quantum effects might resolve the black hole information loss problem.\\

\textbf{EGB gravitational theory:}  
We begin by exploring the $D$-dimensional EGB theory---an elegant extension of Einstein's gravity by Lanczos \cite{Lanczos:1938sf} and Lovelock \cite{Lovelock:1971yv, Lovelock:1972vz}---through its action minimally coupled to matter fields and formulated in the absence of a cosmological constant in $D$-dimensional spacetime:
\begin{eqnarray}\label{1}
\mathcal{S} = \frac{1}{2 k^2_{D} } \int d^{D}x \sqrt{-g} \left[R + \alpha \mathcal{L}_{GB}\right].
\end{eqnarray}
Here, $k_{D} \equiv \sqrt{8\pi G_{D}}$ (where $G_{D}$ is the $D$-dimensional gravitational constant), $\alpha$ is the GB coupling constant---having dimensions of length squared and related to the string tension in string theory---$g$ denotes the determinant of the metric $g_{\mu\nu}$, $R$ is the Ricci scalar, and the GB Lagrangian density $\mathcal{L}_{GB}$ is given by
\begin{equation}\label{GBLag}
\mathcal{L}_{GB}=R^{\mu\nu\rho\sigma}R_{\mu\nu\rho\sigma}-4 R^{\mu\nu}R_{\nu\mu}+R^2.
\end{equation}
Here, $R$ denotes the Ricci scalar, $R_{\mu\nu}$ the Ricci tensor, and $R_{\mu\nu\rho\sigma}$ the Riemann tensor. The gravitational field equations follow from varying the action~(\ref{1}) with respect to $g_{\mu\nu} $~\cite{Lanczos:1938sf}:
\begin{eqnarray}\label{eq:fieldEGB}
R_{\mu\nu} -\frac{1}{2}Rg_{\mu\nu} + \alpha H_{\mu\nu} = T_{\mu\nu},
\end{eqnarray}
where the Lanczos tensor is given by
\begin{eqnarray}
H_{\mu\nu} &=&2\left(RR^{\mu\nu} -2R_{\mu\sigma}R^{\sigma}{\nu}-2R_{\mu\sigma\nu\rho}R^{\sigma\rho} -R_{\mu\sigma\nu\delta}R^{\sigma\rho\delta}_{\nu}\right)\nonumber\\&&-\frac{1}{2}\mathcal{L}_{GB}g_{\mu\nu}.\n
\end{eqnarray}
In $D>4$, the GB term modifies the dynamics while keeping the equations of motion second-order. In exactly four dimensions, however, $\mathcal{L}_{GB}$ reduces to a topological invariant and does not affect the field equations unless the coupling is rescaled as $\alpha \rightarrow \alpha/(D-4)$ before taking the limit $D \to 4$, as proposed in~\cite{Glavan:2019inb}.

To seek a static, spherically symmetric vacuum solution, we begin with the general metric ansatz
\begin{equation}
ds^2 = -h(r) \, dt^2 + \frac{dr^2}{f(r)} + r^2 d\Omega_{D-2}^2,
\label{eq:ansatz}
\end{equation}
where $d\Omega_{D-2}^2$ is the metric on the unit $(D-2)$-sphere. The $(tr)$ component of~\eqref{eq:fieldEGB} in vacuum implies $f(r) = h(r)$, so the metric simplifies to
\begin{equation}
ds^2 = -h(r) \, dt^2 + \frac{dr^2}{h(r)} + r^2 d\Omega_{D-2}^2.
\label{eq:metric}
\end{equation}

From this line element we compute the curvature quantities. The nonzero Ricci tensor components are
\begin{align}
R_{tt} &= \frac{h}{2} \left[ \frac{d^2 h}{dr^2} + \frac{2}{r}\frac{dh}{dr} \right], \\
R_{rr} &= -\frac{1}{2h} \left[ \frac{d^2 h}{dr^2} + \frac{2}{r}\frac{dh}{dr} \right], \\ 
R_{\theta\theta} &= 1 - h - \frac{r}{2}\frac{dh}{dr} - \frac{r^2 }{2}\frac{d^2 h}{dr^2}, \\
R_{\phi\phi} &= \sin^2\theta \, R_{\theta\theta},
\end{align}
and the Ricci scalar reads
\begin{equation}
R = -\frac{d^2 h}{dr^2} - \frac{4}{r}\frac{dh}{dr} - \frac{2[h-1]}{r^2}.
\end{equation}
The Einstein tensor components follow as
\begin{align}
G^t{}_t &= \frac{1}{r}\frac{dh}{dr} + \frac{h-1}{r^2}, \\
G^r{}_r &= \frac{1}{r}\frac{dh}{dr} + \frac{h-1}{r^2}, \\
G^\theta{}_\theta &= \frac{1}{2}\frac{d^2 h}{dr^2} + \frac{1}{r}\frac{dh}{dr}.
\end{align}

The GB invariant for~\eqref{eq:metric} is
\begin{equation}
\mathcal{L}_{GB} = \frac{2(h-1)}{r^2}\frac{d^2 h}{dr^2} + \frac{2}{r^2}\left(\frac{dh}{dr}\right)^2 + \frac{4(h-1) }{r^3}\frac{dh}{dr},
\end{equation}
leading to the Lanczos tensor components
\begin{align}
H^t{}_t &= \frac{2(h-1)}{r^3}\frac{dh}{dr}, \\
H^r{}_r &= \frac{2(h-1)}{r^3}\frac{dh}{dr}, \\
H^\theta{}_\theta &= \frac{(h-1)^2}{r^4}.
\end{align}

Substituting these expressions into~\eqref{eq:fieldEGB} in vacuum ($T_{\mu\nu}=0$) yields a single independent equation
\begin{equation}
\left(r^2+ \alpha (h-1)\right) \frac{dh}{dr} + r(h-1)  = 0.
\label{eq:master}
\end{equation}

It is convenient to define an auxiliary function $\chi(r) \equiv 1 - h(r)$, for which Eq.~\eqref{eq:master} becomes
\begin{equation}
\left( r^2 + \alpha \chi \right)\frac{d\chi}{dr} + r \chi = 0.
\end{equation}
Integrating in $D$ dimensions gives
\begin{equation}
r^{D-3} \chi + \tilde{\alpha} \, r^{D-5} \chi^2 = \mu,
\end{equation}
with $\tilde{\alpha} = (D-3)(D-4)\alpha$ and $\mu$ an integration constant related to the ADM mass \footnote{The ADM mass (after Arnowitt, Deser, and Misner) represents the total mass--energy of an isolated system in an asymptotically flat spacetime. It is determined from the $1/r$ behavior of the metric at large distances and accounts for matter, fields, and the gravitational binding energy.}.

Solving this quadratic equation for $\chi(r)$ and restoring $h(r) = 1 - \chi(r)$ gives the Boulware--Deser solution~\cite{Boulware:1985wk}:
\begin{equation}
h(r) = 1 + \frac{r^2}{2\tilde{\alpha}} \left[ 1 - \sqrt{1 + \frac{4\tilde{\alpha} \mu}{r^{D-1}}} \right].
\end{equation}
Finally, taking the $D \to 4$ limit following the Glavan--Lin prescription $\alpha \rightarrow \alpha/(D-4)$ leads to the uncharged four-dimensional EGB black hole metric,
\begin{equation}
h(r) = 1 + \frac{r^2}{2\alpha} \left[ 1 - \sqrt{ 1 + \frac{8\alpha M}{r^{3}} } \right],
\end{equation}
where $M$ is the ADM mass of the black hole. This solution follows directly from integrating the field equations~\eqref{eq:fieldEGB} in the vacuum case ($T_{\mu\nu}=0$) for the static, spherically symmetric ansatz~\eqref{eq:metric}.\\

\textbf{Dymnikova's profile and GUP correction:} Building on the work of Dymnikova and Ansoldi~\cite{Dymnikova:1996plb,Ansoldi:2008jw}, the energy density profile in four dimensions associated with the Dymnikova vacuum~\cite{Dymnikova:1992ux} can be seen as a gravitational counterpart to the Schwinger effect. In quantum electrodynamics, this effect describes how a strong, uniform electric field produces electron-positron pairs at a rate $\gamma \sim e^{-E_{\textit{crit}}/E}$, where vacuum polarization occurs once the field exceeds the critical strength $E_{\textit{crit}} = \frac{\pi \hbar m_e^2}{e}$, with $m_e$ and $e$ denoting the electron mass and charge.

By analogy, the gravitational equivalent arises by relating the electric field to gravitational tension via curvature, leading to
\begin{small}
\begin{equation}\label{identification}
E \sim \frac{1}{r^3}, \quad \frac{E_{\textit{crit}}}{E} = \frac{r^3}{a_1 a_2^2} \ ,
\end{equation}
\end{small}
where $a_1 = 2M$ and $a_2$ characterize the curvature of the de Sitter core. This analogy gives rise to the four-dimensional \textit{Dymnikova} density profile, expressed as $\rho(r)=\rho_s e^{\left[-\frac{r^3}{a^3}\right]}$. We define $a^3 = a_1 a_2^2$, with $a_2^2 = \frac{3}{8\pi \rho_s}$ to ensure a de Sitter core. In~\cite{Haouat:2013yba, Ong:2020tvo}, the Schwinger effect was corrected using a minimal length $l$ from the GUP, where $\beta = l^2$. Note that in this study we treat $\beta$ as a phenomenological parameter with macroscopic values (km$^2$), used primarily to explore and highlight the qualitative impact of GUP-inspired corrections on astrophysical black holes. For further discussions on the upper bounds of $\beta$, see Refs.~\cite{Neves:2019lio, Jusufi:2020wmp}. For small $\beta$, the pair production rate becomes $\gamma \sim e^{\frac{1}{E}[E^2 A(\beta) - B]}$, with $A(\beta)$ and $B$ depending on the electron's mass and charge. The GUP is given by
\begin{small}
\begin{equation}
\Delta X\Delta P\sim \frac{\hbar}{2}\left[1+\frac{\beta}{\hbar}(\Delta P)^2\right].
\end{equation}
\end{small}
To capture GUP effects in particle production, Haouat et al.~\cite{Haouat:2013yba} derived the expression
\begin{small}
\begin{equation}\label{fullproduction}
\gamma \sim \sinh^2\left(\frac{\pi}{a_4} \sqrt{1 - a_3}\right) \cosh^{-2}\left(\frac{\pi}{a_4} \sqrt{1 - \frac{1}{4}a_4^2}\right),
\end{equation}
\end{small}
where $a_3 = \beta m^2$ and $a_4 = \beta e E$. In the small $\beta$ limit ($a_3, a_4 \ll 1$), this simplifies to
\begin{small}
\begin{equation}\label{correctaappdymnikova}
\gamma \sim e^{\frac{\pi}{4a_4}\left[ \beta^2 - 4 a_3 \right]}.
\end{equation}
\end{small}
To align with the results in~\cite{Estrada:2023pny}, we refine earlier identifications: the curvature tension, expressed by the Kretschmann scalar, follows $\sqrt{K} \sim a_1 / r^3$, and $m^2 \sim 1/a_2^2$. This leads to
\begin{equation}\label{correctiden}
\frac{e E}{\pi} \sim \sqrt{K} \sim \frac{a_1}{r^3}, \quad \Rightarrow \quad \frac{E_\textit{crit}}{E} = \frac{r^3}{a_1 a_2^2}.
\end{equation}
This aligns perfectly with Dymnikova's original identification in Eq.~\eqref{identification}, highlighting the importance of carefully tracking how new parameters enter the theory. To smoothly recover the standard Dymnikova density in the limit $\beta \to 0$, we must retain the earlier identifications. By substituting Eq.~\eqref{correctiden} into Eq.~\eqref{correctaappdymnikova}, we arrive at the GUP-Dymnikova-Schwinger density profile~\cite{Estrada:2023pny}, expressed as
\begin{eqnarray}\label{RhoCV}
\rho(r)&=& \rho_s e^{\left[-\frac{r^3}{a^3}+\beta \frac{\delta}{r^3}\right]}  \approx \rho_s e^{\left[-\frac{r^3}{a^3}\right]}\left[1+\beta \frac{\delta}{r^3}\right],
\end{eqnarray}
where~$\delta = \frac{ \pi^2 a_1}{4}$. When $\frac{\beta\delta}{r^3} \ll 1$---a condition satisfied at sufficiently large $r$ for moderate values of $\beta$---the second exponential in Eq.~(\ref{RhoCV}) can be expanded, leading to the asymptotic expression
\begin{small}
\begin{eqnarray}
\rho(r) \ \sim\ \rho_s\,e^{\left[-\frac{r^3}{a^3}\right]} \left[1+\beta\frac{\delta}{r^3}+\beta^2\frac{\delta^2}{2\,r^6}
+\mathcal{O}\!\left(r^{-9}\right)\right],~r \to \infty.~~
\end{eqnarray}
\end{small}
At large distances, the density falls off extremely rapidly, with the leading term exhibiting a super-exponential decay $\rho(r) \propto e^{-r^3/a^3}$ determined by the scale $a$. The $\beta$-dependent term modifies this behaviour only slightly, introducing corrections that decay algebraically as $r^{-3}$. The expansion above remains accurate provided $r \gg a$ and $\beta\delta/r^3 \ll 1$, ensuring fast convergence of the series for $e^{\beta\delta/r^3}$. If these conditions are not met---for example, if $\beta$ is large enough that $\beta\delta/r^3$ is not small---then the truncated series is no longer reliable and the full factorised form $\rho_s e^{-r^3/a^3} e^{\beta\delta/r^3}$ should be retained.

When a matter source is included, the $tt$ component of the gravitational field equations becomes
\begin{eqnarray}\label{5}
    T^t_t=-\rho(r)&=& \left[\frac{1}{r} - \frac{ 2\alpha(h-1)}{r^3}\right]\frac{dh}{dr} +\frac{ \alpha(h-1)^2}{r^4}\n\\&&\frac{1}{r^2}\left[h-1\right].~~
\end{eqnarray}
Introducing $\chi(r)\equiv h(r)-1$, Eq.~\eqref{5} reduces to a quadratic structure that can be written as a total derivative,
\begin{eqnarray}
\frac{d}{dr}\!\left(r^{3}\chi+\alpha \chi^{2}\right)=-\frac{1}{\pi}\,\frac{d\eta(r)}{dr},
\end{eqnarray}
where $\eta(r)=4\pi\int^r \epsilon^{2}\rho(\epsilon)\,d\epsilon$. Integrating yields the algebraic relation
\begin{eqnarray}
\alpha \chi^{2}+r^{3}\chi-\left(\eta_{m}-\tfrac{1}{\pi}\eta(r)\right)=0,
\end{eqnarray}
with $\eta_m$ an integration constant. Solving this quadratic gives the two branches
\begin{eqnarray}\label{6}
h_{\pm}(r)=1+\frac{r^{2}}{2\alpha}\left[1\pm\sqrt{1-\frac{4\alpha}{r^{3}}\eta_m+\frac{\alpha}{\pi r^{3}}\eta(r)}\;\right],
\end{eqnarray}
with 
\begin{eqnarray}\label{6a}
   \eta(r) &=& \frac{4}{3} \pi  \rho_s \left[\beta  \delta Ei\left[-\frac{r^3}{a^3}\right]-a^3 e^{-\frac{r^3}{a^3}}\right],
\end{eqnarray}
where $Ei[z]$ being the exponential integral function. The physically relevant solution is the minus branch,
\begin{eqnarray}\label{7}
h_{-}(r)=1+\frac{r^{2}}{2\alpha}\left[1-\sqrt{1-\frac{4\alpha}{r^{3}}\eta_m+\frac{\alpha}{\pi r^{3}}\eta(r)}\;\right],
\end{eqnarray}
with $\eta_m=-2M$, ensuring that the Schwarzschild limit is recovered for $\alpha,\rho(r)\to0$.

\begin{figure*}[ht]
\centering
\includegraphics[width=4.35cm,height=5.35cm]{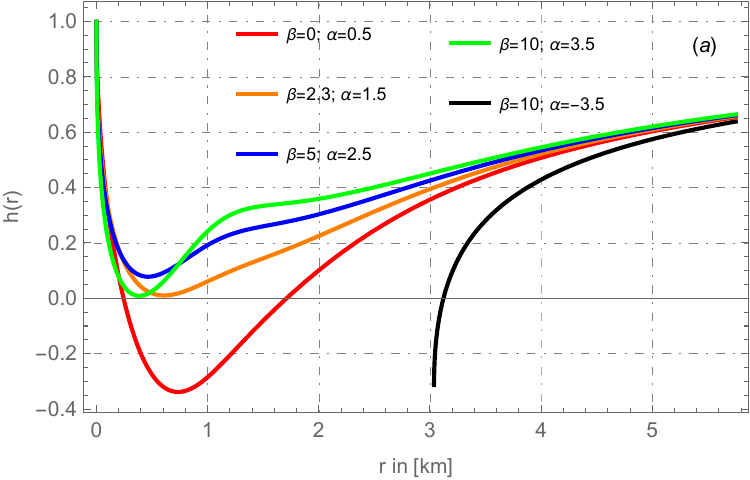}
\includegraphics[width=4.35cm,height=5.35cm]{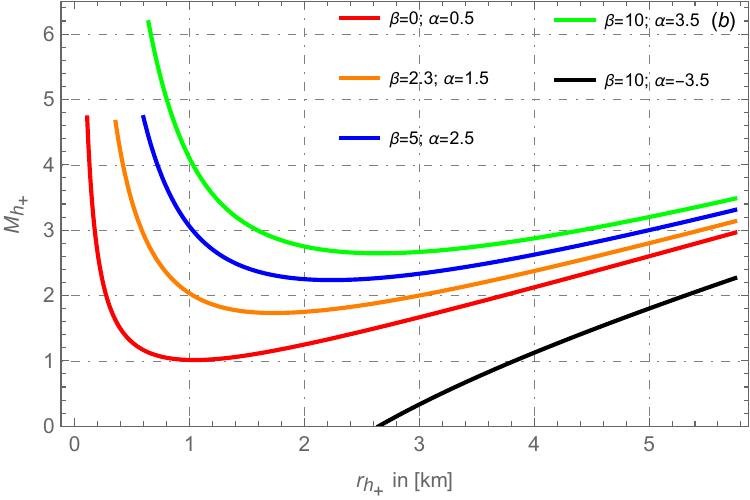}
\includegraphics[width=4.35cm,height=5.35cm]{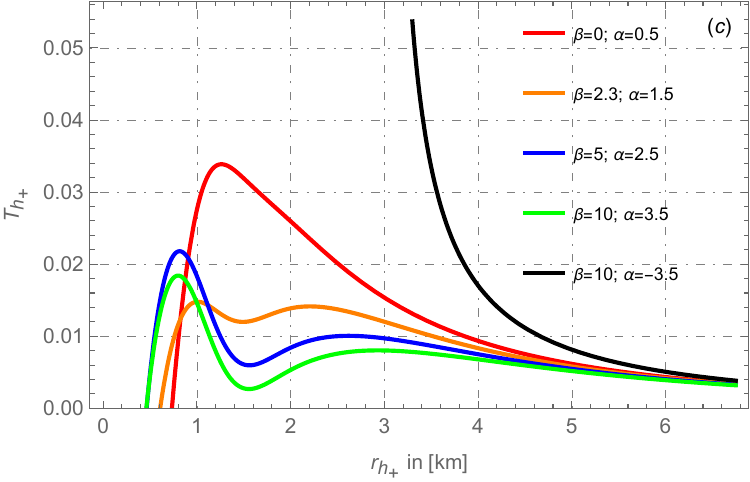}
\includegraphics[width=4.35cm,height=5.35cm]{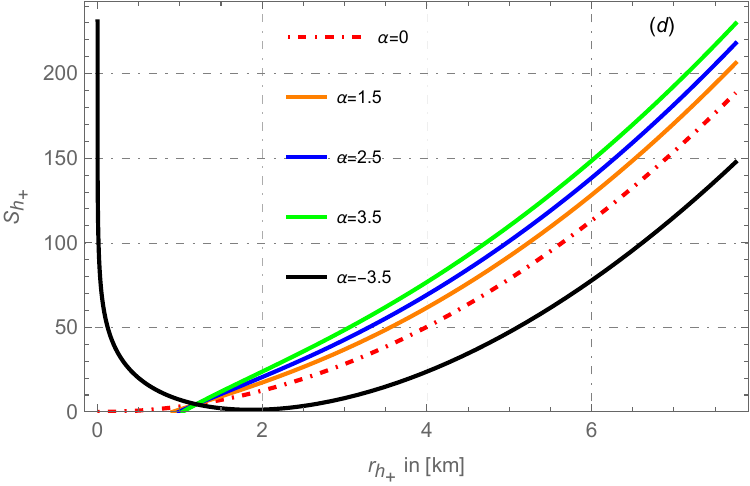}
\caption{ The figure panels demonstrate how variations in the GUP parameter $\beta$ (in $km^2$) and GB coupling constant $\alpha$ affect key thermodynamic and dynamical quantities: the metric function $h(r)$---(a), black hole mass $M_{{h_+}}$---(b), Hawking temperature $T_{H_{h_+}}$---(c), and entropy $S_{{h_+}}$---(d). These results, obtained under the parameter set $\eta_m = -1$ $km$, $\delta = 1$ $km$, $a = 1$ $km$, and $\rho_s = 0.9$ $km^{-2}$, highlight the significant role of quantum corrections in shaping black hole behavior.
}\label{fig1}
\end{figure*}

For $\rho(r)$ with $\eta(r) = 0$, Eq.~\eqref{7} matches the result in~\cite{Guo:2020zmf}. The conditions $tt = rr$ and ${\theta}{\theta} = {\phi}{\phi}$ follow from the EGB equations. The ${\theta}{\theta}$ component reads:
\begin{eqnarray}\label{8}
    P_t&=&\frac{1}{r}\left[ 1 - \frac{2\alpha(h-1)}{r^2}\right]\frac{dh}{dr} + \frac{1}{2}\left[1- \frac{2\alpha}{r^2}- \frac{\alpha^2 h}{r^2}\right]\frac{d^2h}{dr^2} \n \\&&+ \frac{\alpha}{r^2}\left(\frac{dh}{dr}\right)^2  + \frac{\alpha(h-1)^2}{r^4},
\end{eqnarray}
where ${P_t} = T^{\theta}_{\theta} = T^{\phi}_{\phi}$. After applying Eq.~\eqref{5} and simplifying, we find: 
\begin{eqnarray}\label{9}
   \frac{r}{2}\frac{d\rho}{dr}+ \rho +{P_t}=0.
\end{eqnarray}
This is equivalent to $\nabla_\mu T^\mu_r = 0$, confirming that Eq.~\eqref{7} holds without requiring additional physics. In Fig.~\ref{fig1}---(a), we plot the lapse function $h(r)$---which characterizes the spacetime geometry---as a function of radial distance, based on Eq.~(\ref{7}). The plot considers various values of $\alpha = 0.5, 1.5, 2.5, 3.5, -3.5$ with corresponding $\beta = 0, 2.3, 5, 10, 10$ $km^2$, while keeping $\eta_m = -1$ $km$, $\delta = 1$ $km$, $a = 1$ $km$, and $\rho_s = 0.9$ $km^{-2}$ fixed. The key feature is where $h(r)$ intersects the horizontal axis, indicating the \textit{Cauchy horizon} ($r_c$) and the \textit{event horizon} ($r_{h_+}$). Two distinct horizons appear for $\alpha < 1.5$ with $\beta < 2.3$ $km^2$, and for $\alpha > 3.5$ with $\beta > 10$ $km^2$, representing standard black hole structures. For negative $\alpha$ and any positive $\beta$, only a single horizon exists, suggesting a different causal structure. At critical values---$\alpha = 1.5$, $\beta = 2.3$ $km^2$ and $\alpha = 3.5$, $\beta = 10$ $km^2$---the horizons coincide, forming an \textit{extremal black hole} with a degenerate horizon. Beyond these ranges, $h(r)$ remains nonzero, implying a regular spacetime without black holes. Additionally, we observe that increasing $\alpha$ and $\beta$ leads to a smaller event horizon, indicating a more compact black hole.\\

\textbf{Thermodynamic perspective on quantum corrections to Dymnikova-Schwinger black holes:}
The gravitational mass of the black hole can be obtained by solving $h(r_{h_+}) = 0$, yielding:
\begin{small}
\begin{eqnarray}\label{16.2}
M_{h_+} = \frac{1}{2}\left[r_{h_+}+\frac{\alpha}{r_{h_+}} - \frac{\rho_s}{12}   \left[\delta\beta   Ei\left[-\frac{r_{h_+}^3}{a^3}\right]-a^3 e^{-\frac{r_{h_+}^3}{a^3}}\right]\right].~~
\end{eqnarray}
\end{small}
Fig.~\ref{fig1}---(b) illustrates how the black hole mass $M_{h_+}$ changes with the horizon radius $r_{h_+}$ when quantum corrections (through $\beta$ ( in $km^2$)) and the GB coupling ($\alpha$) are taken into account. A minimum mass, $M_{h_+}^{\text{min}}$, appears at small $r_{h_+}$, marking the point where the black hole becomes extremal, with a single, degenerate horizon at $r_e$. Notably, the extremal black hole is heavier for $\alpha = 3.5$, $\beta = 10$ $km^2$ than for $\alpha = 1.5$, $\beta = 2.3$ $km^2$, showing how these parameters significantly shape the black hole's structure. To investigate how black holes behave as they evaporate, we consider the Hawking temperature~\cite{Wald:1984rg}, given by~\cite{Hawking:1975vcx}:
\begin{eqnarray}
  T_H = \frac{1}{4\pi}\sqrt{-\frac{1}{2}\nabla_{\mu} \xi_{\mu} \nabla^{\mu} \xi^{\mu}},  
\end{eqnarray}
The surface gravity-temperature relation explains black hole radiation and evaporation. Using the time-like Killing vector $\xi^\mu = \partial/\partial t$ from metric~(\ref{eq:metric}), the Hawking temperature is~\cite{Wald:1984rg}:
\begin{eqnarray}\label{17.2}
    T_H(r_{h_+}) = \frac{1}{4\pi}\frac{dh(r_{h_+})}{dr}.
\end{eqnarray}
At the horizon $r_{h_+}$, where $h(r_{h_+}) = 0$, it becomes:
\begin{eqnarray}\label{18.2}
T_H(r_{h_+}) &=& \frac{r_{h_+}}{4\pi\left({r_{h_+}^2 + 2\alpha}\right)}\Bigg[1 - \frac{\alpha}{r_{h_+}^2} - \frac{r_{h_+}^2}{4} \rho_s e^{\left[-\frac{r_{h_+}^3}{a^3}\right]}\n\\&&\left[1+\beta \frac{\delta}{r_{h_+}^3}\right]~\Bigg].
\end{eqnarray}
For $\alpha \to 0$ and $\rho = 0$, this reduces to the Schwarzschild temperature $T_H = 1/(4\pi r_{h_+})$~\cite{Guo:2020zmf}. Fig.~\ref{fig1}---(c) shows the Hawking temperature $T_{H_{h_+}}$ of Dymnikova-Schwinger black holes versus horizon radius $r_{h_+}$. The temperature drops to zero when $r_{h_+}$ satisfies:
\begin{eqnarray}
 4r_{h_+}^2 - r_{h_+}^4 \rho_s e^{\left[-\frac{r_{h_+}^3}{a^3}\right]}\left[1+\beta \frac{\delta}{r_{h_+}^3}\right] -4\alpha = 0,
\end{eqnarray}
Increasing $\alpha$ and $\beta$ lowers the peak Hawking temperature, as seen in Fig.~\ref{fig1}---(c). At the end of evaporation, black holes become stable, zero-temperature remnants at radius $r_e$. Notably, the remnant with $\alpha=3.5$, $\beta=10$ $km^2$ is slightly smaller than that with $\alpha=1.5$, $\beta=2.3$ $km^2$.

Using the first law of black hole thermodynamics
\begin{equation}
dM(r_{h_+})=T_H(r_{h_+})dS,
\label{24}
\end{equation}
the entropy can be obtained by integrating:
\begin{equation}
S=\int \frac{dM(r_{h_+})}{T_H(r_{h_+})}=\int \frac{1}{T_H(r_{h_+})}\frac{\partial M(r_{h_+})}{\partial r_{h_+}}dr_{h_+}.
\label{25}
\end{equation}
Substituting Eqs.~(\ref{16.2}) and (\ref{18.2}), we find:
\begin{eqnarray}\label{20.2}
    S_{h_+} = \frac{A_{BH}}{4} + 2\pi \alpha \ln\left[\frac{A_{BH}}{A_0}\right],
\end{eqnarray}
where $A_{BH} = 4\pi r_{h_+}^2$ and $A_{0}$ is a constant reference area \cite{Fernandes:2020nbq}. The first term reproduces the Bekenstein--Hawking area law, while the second represents a purely geometric higher-curvature correction from the Gauss--Bonnet term \cite{Cai:2009ua, Cai:2014jea}, reflecting extra classical degrees of freedom and persisting even without quantum effects. By contrast, GUP-based models predict a similar $\ln A_{BH}$ term,
\begin{equation}
S_{\mathrm{GUP}} \simeq \frac{A_{BH}}{4} + \beta \ln\left(\frac{A_{BH}}{A_P}\right) + \cdots,
\end{equation}
but originating from Planck-scale modifications to the uncertainty principle.  
Both mechanisms can contribute additively only if the GB term is treated purely as a classical correction to GR and the GUP term as a distinct quantum effect. In the Dymnikova--Schwinger black hole case \cite{Guo:2020zmf}, quantum corrections are absent, and setting $\alpha = 0$ in Eq.~(\ref{20.2}) recovers the standard Bekenstein--Hawking entropy, as clearly illustrated by the red dot-dashed line in Fig.~\ref{fig1}---(d). For black holes with large horizon radii $r_{h_+}$, the logarithmic correction is negligible, reaffirming the dominance of the classical area law. However, for smaller black holes, this correction becomes a crucial component of the entropy, and its omission would lead to a fundamentally incomplete description of the thermodynamics.

\begin{figure*}[ht]
\centering
\includegraphics[width=4.35cm,height=5.35cm]{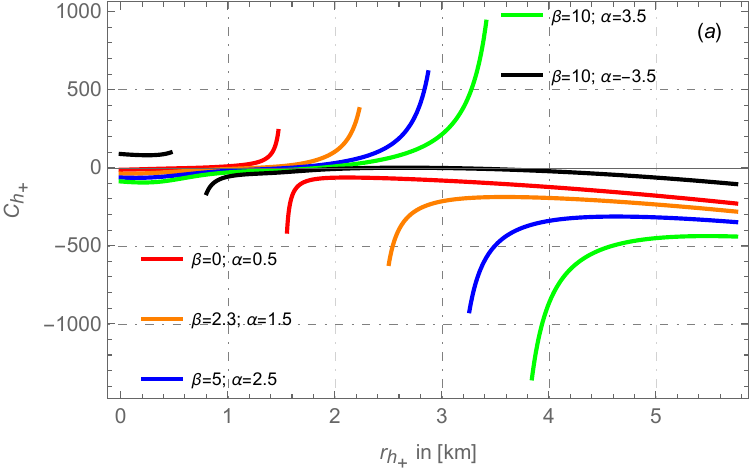}
\includegraphics[width=4.35cm,height=5.35cm]{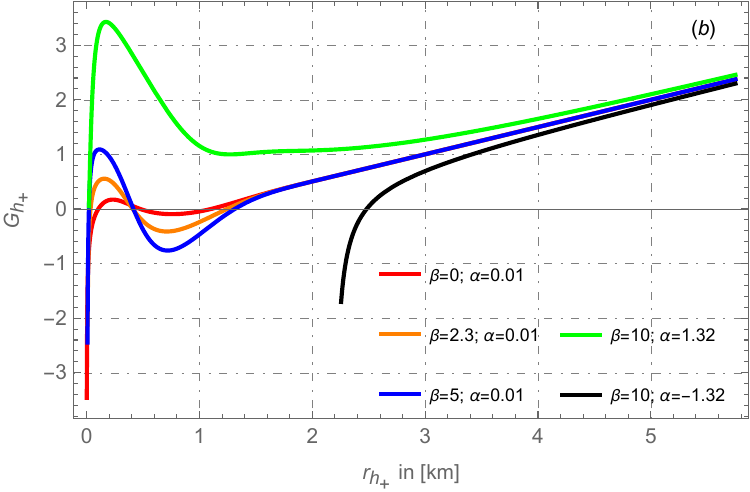}
\includegraphics[width=4.35cm,height=5.35cm]{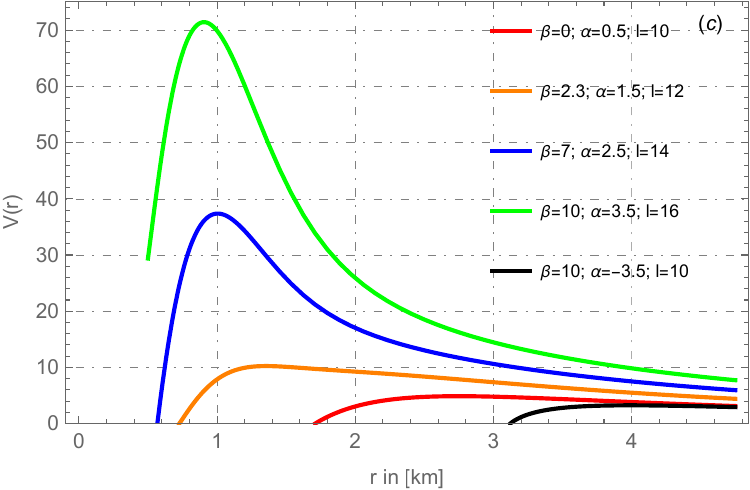}
\includegraphics[width=4.35cm,height=5.35cm]{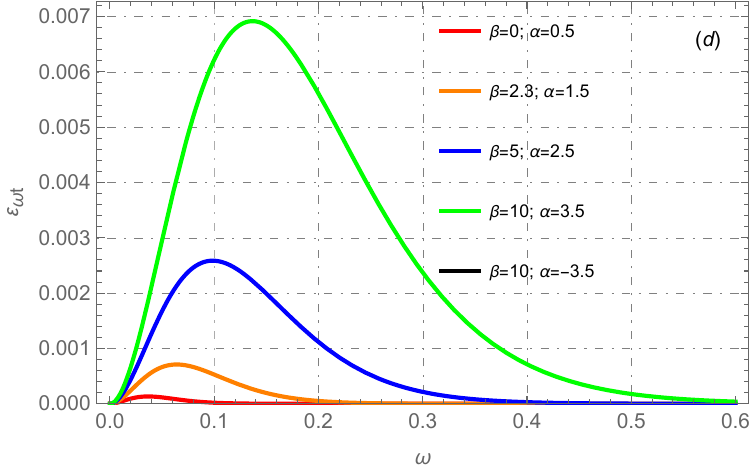}
\caption{ The figure panels illustrate the influence of varying the GUP parameter $\beta$ (in $km^2$) and GB coupling constant $\alpha$ on critical thermodynamic and dynamical properties: specific heat $C_{h_+}$---(a), Gibbs free energy $G_{h_+}$---(b), effective potential $V(r_{h_+})$---(c), and energy emission $\mathcal{E}_{\omega t}$---(d). These results, derived using the parameter set $\eta_m = -1$ $km$, $\delta = 1$ $km$, $a = 1$ $km$, and $\rho_s = 0.9$ $km^{-2}$, underscore the substantial role of quantum corrections in determining black hole behavior.}\label{fig2}
\end{figure*}

We use Eq. (\ref{25}) to find the heat capacity, which helps us understand the black hole's stability
\begin{small}
\begin{equation}
C_h(r_{h_+})=T_H\left(\frac{\partial S}{\partial T_H}\right)_q=\frac{\partial M(r_{h_+})}{\partial T_H(r_{h_+})}=\frac{\partial M(r_{h_+})/\partial r_{h_+}}{\partial T_H(r_{h_+})/\partial r_{h_+}}.
\label{26}
\end{equation}
\end{small}
By applying Eqs. (\ref{16.2}) and (\ref{18.2}), we can uncover
\begin{small}
\begin{equation}
\frac{\partial M(r_{h_+})}{\partial r_{h_+}}= \frac{1}{2}\left[1
-\frac{\alpha}{r_{h_+}^2} - \frac{r_{h_+}^2 \rho_s}{4}  e^{\left[-\frac{r_{h_+}^3}{a^3}\right]}\left[1+\beta \frac{\delta}{r_{h_+}^3}\right]\right].
\label{27}
\end{equation}
\end{small}
\begin{small}
\begin{eqnarray}
\frac{\partial T_H(r_{h_+})}{\partial r_{h_+}}&=&-\frac{1}{16 \pi  \left(r^3+2 \alpha  r_{h_+}\right)^2}\Bigg[-4 \alpha ^2 + 4 r_{h_+}^4 \n\\&&-14 \alpha  r_{h_+}^2 + 3 \rho_s\left(r_{h_+}^7+2 \alpha  r_{h_+}^5\right) e^{-\frac{r_{h_+}^3}{a^3}}\n\\&&\left[ -\frac{ \beta  \delta }{r_{h_+}^4}-\frac{ r_{h_+}^2 }{a^3}\left(\frac{\beta  \delta}{r^3}+1\right)\right] \nonumber\\ &&+ \left(r_{h_+}^6+6 \alpha  r_{h_+}^4\right) \rho_s e^{\left[-\frac{r_{h_+}^3}{a^3}\right]}\left[1+\beta \frac{\delta}{r_{h_+}^3}\right]\Bigg],~~~~~~
\label{28}
\end{eqnarray}
\end{small}
leading to
\begin{small}
\begin{eqnarray}\label{21.2}
    C_h &=& 2 \pi  \left[2 \alpha +r_{h_+}^2\right]^2 \left[4 \alpha +r_{h_+}^4 \rho_s e^{\left[-\frac{r^3}{a^3}\right]}\left[1+\beta \frac{\delta}{r^3}\right]-4 r_{h_+}^2\right]\n\\&&\Bigg[-4 \alpha ^2+4 r_{h_+}^4-14 \alpha  r_{h_+}^2+\left[r_{h_+}^7+2 \alpha  r_{h_+}^5\right] \n\\&&\frac{-3 \rho_s e^{-\frac{r_{h_+}^3}{a^3}} \left(a^3 \beta  \delta+\beta  \delta r_{h_+}^3+ r_{h_+}^6\right)}{a^3 r_{h_+}^4} \n\\&&+\left[r_{h_+}^6+6 \alpha  r_{h_+}^4\right] \rho_s e^{\left[-\frac{r_{h_+}^3}{a^3}\right]}\left[1+\beta \frac{\delta}{r_{h_+}^3}\right]\Bigg]^{-1}.~~~~~
\end{eqnarray}
\end{small}
From Eq.(\ref{26}), the heat capacity diverges when the Hawking temperature reaches an extremum, where $\partial T_H(r_{h_+})/\partial r_{h_+} = 0$ (see Fig.\ref{fig2}---(a)). At a critical radius $r_c$, both the temperature and heat capacity vanish, signaling a first-order phase transition and the formation of a remnant with non-zero mass. Another key point, $r_a$, also satisfies this extremum condition and corresponds to a discontinuity in heat capacity. Black holes are locally stable for $r_c < r_{h_+} < r_a$, but become unstable and begin to evaporate when $r_{h_+} > r_a$.

Finally, we compute the Gibbs free energy, given by:
\begin{small}
\begin{eqnarray}
G_{h_+} = M_{h_+} - T_{h_+}S_{h_+}.    
\end{eqnarray}
\end{small}
By combining Eqs.~(\ref{16.2}), (\ref{18.2}), and (\ref{20.2}), the Gibbs free energy can be readily obtained as:
\begin{small}
\begin{eqnarray}\label{22.2}
G_{h_+} &=&  \frac{\alpha }{r_{h_+}}+\frac{r_{h_+}}{2} -\frac{\rho_s}{24 }   \left[\beta  \delta Ei\left[-\frac{r_{h_+}^3}{a^3}\right]-a^3 e^{-\frac{r_{h_+}^3}{a^3}}\right]-\n\\&&   \Bigg[r_{h_+} \Bigg[-\frac{\alpha }{2 r_{h_+}^2} -\frac{r_{h_+}^2 \rho_s}{4} e^{\left[-\frac{r^3}{a^3}\right]}\left[1+\beta \frac{\delta}{r^3}\right]+1\Bigg]\n\\&&  \Bigg[\pi  r_{h_+}^2+2 \pi  \alpha  \ln \left(\frac{r_{h_+}}{r_{0}}\right)\Bigg]\Bigg]\Bigg[4 \pi  \left(2 \alpha +r_{h_+}^2\right)\Bigg]^{-1}.~~
\end{eqnarray}
\end{small}
\\

\begin{figure*}[ht]
\centering
\includegraphics[width=4.35cm,height=5.35cm]{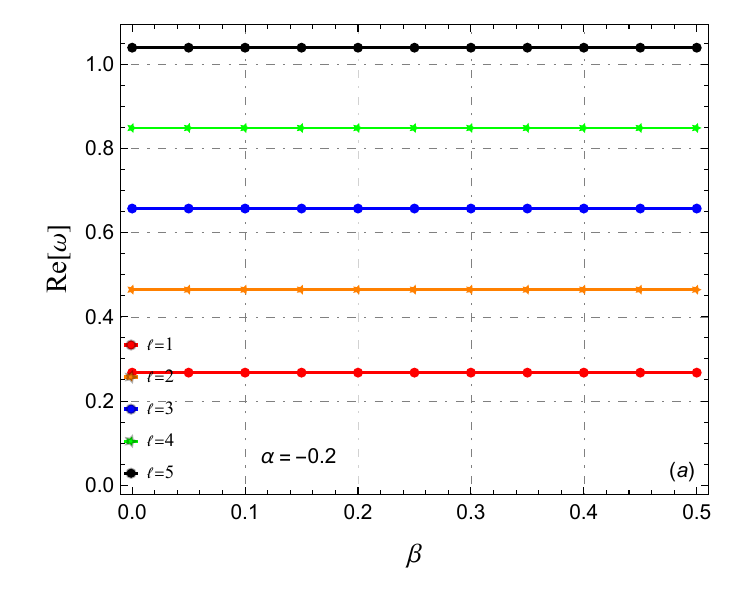}
\includegraphics[width=4.35cm,height=5.35cm]{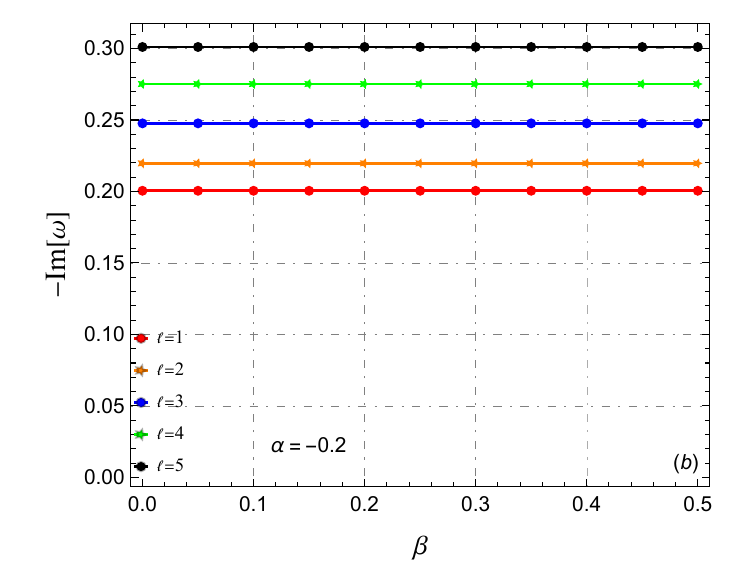}
\includegraphics[width=4.35cm,height=5.35cm]{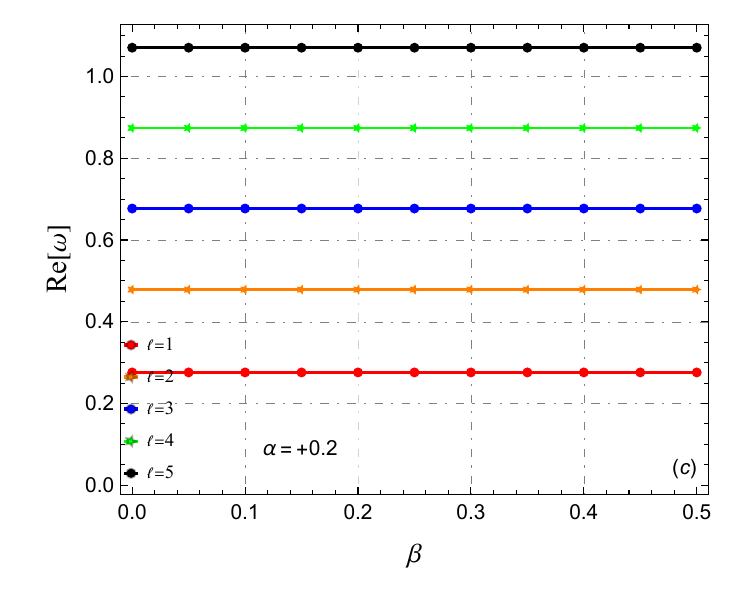}
\includegraphics[width=4.35cm,height=5.35cm]{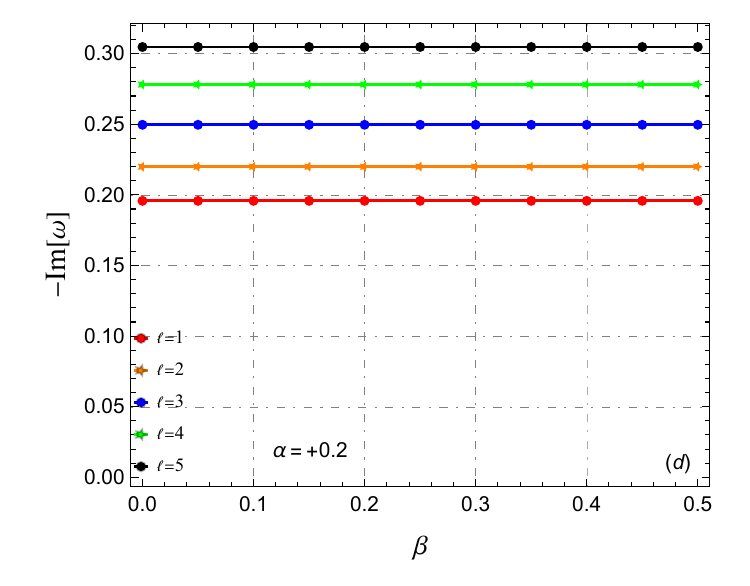}
\caption{ The figure panels illustrate the influence of the GUP parameter $\beta$ (in $km^2$), the GB coupling constant $\alpha$, and the orbital angular momentum $\ell$ on quasinormal mode frequencies $\omega = \omega_{Re} - i\,\omega_{\mathrm{Im}}$: real part $Re[\omega]$---(a) $\alpha = -0.2$, imaginary part $\mathrm{Im}[\omega]$---(b) $\alpha = -0.2$, real part $Re[\omega]$---(c) $\alpha = +0.2$, and imaginary part $\mathrm{Im}[\omega]$---(d) $\alpha = +0.2$. These results, derived using the parameter set $\eta_m = -1$ $km$, $\delta = 1$ $km$, $a = 1$ $km$, and $\rho_s = 0.9$ $km^{-2}$, underscore the substantial role of quantum corrections in determining black hole behavior.}\label{fig3}
\end{figure*}

\textbf{Quasinormal modes:}
Quasinormal modes are the fading echoes of black holes, revealing their mass, charge, and shape. Their complex frequencies capture how these cosmic objects vibrate and settle, offering deep insight into gravity and spacetime \cite{Cai:2014znn, Kubiznak:2016qmn, Konoplya:2002ky}.
\begin{eqnarray}\label{eq18}
\frac{1}{|g|}\partial_{\mu}\left(\sqrt{|g|}\partial^{\mu}\psi\right)=0.
\end{eqnarray}
The Klein-Gordon equation, which describes scalar field perturbations, can be simplified into a radial form:
\begin{small}
\begin{eqnarray}\label{eq19}
h(r)^2 \psi^{''}(r) + h^{'}(r)h(r) \psi^{'}(r)+\left(\omega^2-V\right)\psi(r)=0.
\end{eqnarray}
\end{small}
where primes represent derivatives with respect to $r$, and $h(r)$ relates to the spacetime geometry. By changing to the tortoise coordinate, defined as
\begin{eqnarray}\label{eq20}
dr_{\star}=\pm\frac{dr}{h(r)}.
\end{eqnarray}
The perturbation equation takes a wave-like form:
\begin{eqnarray}\label{eq21}
\frac{d^2\psi}{dr_{\star}} +\left(\omega^2-V\right)\psi=0.
\end{eqnarray}
The effective potential $V(r)$ is given by
\begin{equation}\label{eq22}
V(r) = h(r) \left(  \frac{h'(r)}{r} + \frac{\ell(\ell+1)}{r^2}\right),
\end{equation}
This approach reveals how scalar waves evolve near black holes, with quasinormal modes defined as solutions to Eq.~\eqref{eq21} satisfying $\psi \approx  e^{\pm i\omega r_{\star}}$ as $r_{\star}\rightarrow \pm \infty$ \cite{Konoplya:2005et, Konoplya:2006rv}. Quasinormal modes feature only outgoing waves, representing the black hole's natural ringing after perturbations. The frequency is written as
\begin{eqnarray}\label{eq23a}
\omega =\omega_{Re} - i\omega_{\mathrm{Im}}. 
\end{eqnarray}
where $\omega_{Re}$ is the oscillation frequency, and $\omega_{\mathrm{Im}}$ the decay rate.  The WKB method \cite{Iyer:1986np} approximates quasinormal modes for large $\ell$ ($\ell$ represents the orbital angular momentum, or spherical harmonic index), matching expansions near the horizon and infinity with the potential peak, giving
\begin{eqnarray}\label{eq24}
\frac{Q_0}{\sqrt{2Q^{''}}}=-i\left(n+\frac{1}{2}\right)+O\left(\frac{1}{\ell}\right),
\end{eqnarray}
with
\begin{eqnarray}\label{eq25}
Q_p&=&\omega^2-V(r=r_p),\n\\\label{eq26}
Q^{''}&=&\frac{d^2Q[r(r_{\star})]}{dr^2_{\star}}\n.
\end{eqnarray}
Here, $n$ labels the overtones. From the analytical solution of Eq.~\eqref{eq24}, the quasinormal frequencies have real and imaginary parts:
\begin{eqnarray}\label{eq27}
\omega^2=\frac{ \ell(\ell + 2)h(r_p)}{r_p^2\sqrt{2V^{''}_p}} -i\left(n+\frac{1}{2}\right)h(r_p)\sqrt{2}\sqrt{V^{''}_p}.
\end{eqnarray}
For black holes, this gives the real part as
\begin{equation}
\omega_{Re} = \frac{\sqrt{\ell(\ell+1) h(r_p)}}{r_p},
\end{equation}
where \( r_p \) denotes the position of the peak of the potential.

The imaginary part, which describes the damping of the modes, depends on the curvature of the potential at \( r_p \) and is expressed as
\begin{equation}
\omega_{\mathrm{Im}} = \left(n + \frac{1}{2}\right) \frac{1}{\sqrt{2}} \sqrt{\left| \frac{V''(r_p)}{\omega_{Re}} \right|},
\end{equation}
with \( n \) labeling the overtone number.

Explicitly differentiating \( V(r) \) twice gives
\begin{align}
V''(r_p) =\ &+h(r_p) \left( \frac{6 \ell(\ell+1)}{r_p^4} - \frac{2 h''(r_p)}{r_p^2} + \frac{4 h'(r_p)}{r_p^3} \right)
\nonumber \\
& + 2 h'(r_p) \left( -\frac{2 \ell(\ell+1)}{r_p^3} + \frac{h''(r_p)}{r_p} - \frac{h'(r_p)}{r_p^2} \right)  
\nonumber \\
& +  h''(r_p) \left( \frac{\ell(\ell+1)}{r_p^2} + \frac{h'(r_p)}{r_p} \right) .
\end{align}

Putting all this together, the damping frequency can be written fully in terms of \( h(r_p) \) and its derivatives as
\begin{equation}
\omega_{\mathrm{Im}} = \left(n + \frac{1}{2}\right) \frac{r_p^{1/2}}{\sqrt{2} \left[\ell(\ell+1) h(r_p)\right]^{1/4}} \sqrt{\left| V''(r_p) \right|}.
\end{equation}
This form facilitates the numerical calculation of the quasi-normal mode spectrum upon exploiting the metric function (\ref{7}). Fig.~\ref{fig2}---(c) demonstrates that for $\alpha \geq 1.5$, $\beta \geq 2.3$ $km^2$, and $\ell \geq 12$, the effective potential $V(r)$ forms a clear barrier, ensuring the stability of quasinormal modes. This is confirmed by the negative imaginary parts of the frequencies ($\mathrm{Im}[\omega] = -\omega_{\mathrm{Im}} < 0$). Increasing $\alpha$, $\beta$, and $\ell$ simultaneously intensifies oscillation frequencies and accelerates the damping of scalar perturbations, highlighting the strong influence of quantum corrections on black hole dynamics. We now turn to the analysis of scalar quasinormal modes as a function of the parameter $\beta$, considering orbital angular momenta $\ell = 1$ to $\ell = 5$ for the fundamental overtone ($n = 0$) and two GB coupling constants, $\alpha = -0.2$ and $\alpha = +0.2$, over the range $\beta \in [0, 0.5]$ $km^2$. Fig.~\ref{fig3} [\,$\mathrm{Re}[\omega]$---(a) $\alpha = -0.2$, $\mathrm{Im}[\omega]$---(b) $\alpha = -0.2$, $\mathrm{Re}[\omega]$---(c) $\alpha = +0.2$, and $\mathrm{Im}[\omega]$---(d) $\alpha = +0.2$\,] shows that the real part of the frequency, $\mathrm{Re}[\omega]$, increases slightly and monotonically with $\beta$ for each $\ell$. Although the variation is subtle, this consistent trend suggests a weak dependence on $\beta$. In contrast, the imaginary part, $\mathrm{Im}[\omega]$, exhibits a systematic decrease in absolute magnitude as $\beta$ grows. This reduction in $|\mathrm{Im}[\omega]|$ indicates weaker damping (or equivalently, a longer decay time) with increasing $\beta$, hinting at reduced dissipative effects in the system. Since all imaginary parts in Table~\ref{tab1} are negative, the modes remain stable. Physically, $\mathrm{Re}[\omega]$ sets the oscillation frequency. Table~\ref{tab1} confirms that increasing $\beta$ and $\ell$ for both values of $\alpha$ leads to a higher oscillation frequency $\mathrm{Re}[\omega]$ and a smaller absolute value of the imaginary part $|\mathrm{Im}[\omega]|$. Consequently, larger $\beta$, $\alpha$, and $\ell$ correspond to scalar perturbations that oscillate more rapidly but decay more slowly.
\\

\begin{table}[!htp]
\centering
\caption{Quasinormal mode frequencies $\omega =\omega_{Re} - i\,\omega_{Im}$ for various parameters $\beta$, $\alpha$, and the orbital angular momentum $\ell$.}\label{tab1}
\begin{minipage}{0.3\textheight}
 \scalebox{01.}{\begin{tabular}{|c||c||c||c|c|} \hline
$\beta$ in $km^2$ & $ \alpha$ & $\ell$ & $Re[\omega ]$ & $ \mathrm{Im}[\omega]$   \\[0.15cm]
\hline
0.05 & -0.2 & 1 & 0.27259 & -0.197637 \\\hline
 0.05 & -0.2 & 2 & 0.472999 & -0.219925 \\\hline
 0.05 & -0.2 & 3 & 0.669099 & -0.248952 \\\hline
 0.05 & -0.2 & 4 & 0.863859 & -0.27707 \\\hline
 0.05 & -0.2 & 5 & 1.05803 & -0.303405 \\\hline
 0.05 & 0.2 & 1 & 0.27259 & -0.197637 \\\hline
 0.05 & 0.2 & 2 & 0.472999 & -0.219925 \\\hline
 0.05 & 0.2 & 3 & 0.669099 & -0.248952 \\\hline
 0.05 & 0.2 & 4 & 0.863859 & -0.27707 \\\hline
 0.05 & 0.2 & 5 & 1.05803 & -0.303405 \\\hline
 0.35 & -0.2 & 1 & 0.279675 & -0.193708 \\\hline
 0.35 & -0.2 & 2 & 0.484821 & -0.219778 \\\hline
 0.35 & -0.2 & 3 & 0.685727 & -0.250149 \\\hline
 0.35 & -0.2 & 4 & 0.885296 & -0.279024 \\\hline
 0.35 & -0.2 & 5 & 1.08427 & -0.305887 \\\hline
 0.35 & 0.2 & 1 & 0.279675 & -0.193708 \\\hline
 0.35 & 0.2 & 2 & 0.484821 & -0.219778 \\\hline
 0.35 & 0.2 & 3 & 0.685727 & -0.250149 \\\hline
 0.35 & 0.2 & 4 & 0.885296 & -0.279024 \\\hline
 0.35 & 0.2 & 5 & 1.08427 & -0.305887 \\ 
\hline
\end{tabular}}
 \end{minipage}
\end{table}

\textbf{ Emission Energy:}
Quantum effects near a black hole constantly produce particles, fueling Hawking radiation and slow evaporation. For distant observers, the emission region aligns with the black hole,s shadow, defined by the cross-section $\sigma_{\text{lim}} \approx \pi r_{sh}^2$ \cite{Wei:2013kza}, where $r_{sh}$ relates to the critical orbit and observer position \cite{Perlick:2021aok}. The energy emission rate is given by \cite{Wei:2013kza, Papnoi:2014aaa, EslamPanah:2020hoj, Yang:2023agi}:
\begin{eqnarray}
    \mathcal{E}_{\omega t}\equiv \frac{d^2 \epsilon}{d\omega dt} = \frac{2 \pi^2 \sigma_{\text{lim}}}{e^{\omega/T_{H_{h_+}}} - 1} \omega^3,
 \end{eqnarray}
The impact of quantum effects and GB coupling on black hole radiation is clearly illustrated in Fig.~\ref{fig2}---(d). When the GB coupling constant $\alpha$ is fixed at a positive value, increasing the quantum correction parameter $\beta$ leads to a significant rise in the energy emission rate $\varepsilon_{\omega t}$. This indicates that under such conditions, the black hole radiates energy more efficiently.

The interplay between the GB term (via $\alpha$) and quantum effects (via $\beta$) boosts the Hawking radiation process, highlighting the important role these modifications play in black hole thermodynamics.\\

\textbf{Concluding remarks:} In this work, we explored a quantum-corrected version of black holes within the framework of four-dimensional EGB gravity, incorporating the Dymnikova-Schwinger energy density profile and corrections from the GUP. By considering the Dymnikova profile as a gravitational analogue of the Schwinger effect, we connected spacetime curvature to quantum particle production and derived a modified energy density that naturally includes quantum corrections via a minimal length scale. Solving the EGB field equations with this matter profile, we found regular black hole solutions---free of singularities---that exhibit rich horizon structures. Depending on the values of the GB coupling ($\alpha$) and GUP parameter ($\beta$), the black holes can have two horizons, a single extremal horizon, or none at all. This shows how quantum corrections and higher-curvature effects fundamentally shape the causal structure of spacetime. From a thermodynamic perspective, we derived expressions for the black hole mass and Hawking temperature. Our results indicate that evaporation halts at a finite radius, leaving behind a stable, zero-temperature remnant. The properties of this remnant---its size, mass, and temperature behavior---are influenced significantly by $\alpha$ and $\beta$, highlighting how modifications from both higher-dimensional gravity and quantum mechanics play a key role in black hole evolution.

In summary, our findings suggest that incorporating quantum effects through GUP and curvature corrections via GB terms can resolve singularities and point toward a consistent picture of black hole remnants---potentially offering a path to new physics beyond GR.

\section*{Acknowledgement}
This research was funded by the Science Committee of the Ministry of Science and Higher Education of the Republic of Kazakhstan (Grant No. AP23487178). We thank the reviewer for their valuable comments and kind suggestions, which have greatly helped to improve the clarity and quality of this work.
 

\end{document}